\documentclass[sigconf]{acmart}

\AtBeginDocument{%
  }

\setcopyright{acmlicensed}
\copyrightyear{2026}
\acmYear{2026}
\acmDOI{XXXXXXX.XXXXXXX}
\acmConference[RecSys‘26]{20th ACM Conference on Recommender Systems}{2026}{Minneapolis, Minnesota, USA}
\acmISBN{978-1-4503-XXXX-X/2018/06}

\usepackage{enumitem}
\usepackage{multirow}
\usepackage{makecell}
\usepackage{algorithm}
\usepackage{algpseudocode}
\usepackage{graphicx}
\usepackage{subcaption}

\begin{document}

\title{From Extraction to Navigation: Progressive Retrieval with Indirectly Infinite Depth}

\author{Linxiao Che}
\authornote{These authors contributed equally to this work.} 
\email{chelinxiao@kuaishou.com}
\affiliation{
\institution{Kuaishou Technology}
\city{Beijing}
\country{China}
}
\author{Shanshan Huang}  
\authornotemark[1]
\email{huangshanshan@kuaishou.com}
\affiliation{%
\institution{Kuaishou Technology}
\city{Beijing}
\country{China}
}

\author{Haitao Lu}
\authornotemark[1]
\email{luhaitao03@kuaishou.com}
\affiliation{%
\institution{Kuaishou Technology}
\city{Beijing}
\country{China}
}

\author{Yijia Sun}
\email{sunyijia@kuaishou.com}
\orcid{0009-0000-8779-2108}
\affiliation{%
\institution{Kuaishou Technology}
\city{Beijing}
\country{China}
}

\author{Qiang Luo}
\authornote{Corresponding authors.} 
\email{luoqiang@kuaishou.com}
\affiliation{
\institution{Kuaishou Technology}
\city{Beijing}
\country{China}
}

\author{Ruiming tang}
\authornotemark[2]
\email{tangruiming@kuaishou.com}
\affiliation{
\institution{Kuaishou Technology}
\city{Beijing}
\country{China}
}

\author{Han Li}
\authornotemark[2]
\email{lihan08@kuaishou.com}
\affiliation{
\institution{Kuaishou Technology}
\city{Beijing}
\country{China}
}

\author{Kun Gai}
\email{gai.kun@qq.com}
\affiliation{%
\institution{Unaffiliated}
\city{Beijing}
\country{China}
}

\author{Guorui Zhou}
\authornotemark[2]
\email{zhouguorui@kuaishou.com}
\affiliation{%
\institution{Kuaishou Technology}
\city{Beijing}
\country{China}
}

\renewcommand{\shortauthors}{Linxiao Che et al.}

\begin{abstract}
Modern large-scale recommender systems are undergoing a paradigm shift in retrieval, moving from static similarity extraction to dynamic navigation. This new paradigm treats retrieval as an iterative, goal-oriented traversal of the complex, high-dimensional item space. Traditional item-to-item (i2i) heuristics are often confined by the ``interest tunnel'' effect, which limits their ability to explore deep-depth interests. Meanwhile, existing structured indexing approaches frequently suffer from ``search drift'' due to their reliance on static entry points and rigid topologies that fail to adapt to evolving real-time intent.

To address these challenges, we propose \textbf{IID-Nav}, a framework that formalizes retrieval as a stateful, autonomous exploration process. IID-Nav introduces three critical innovations: 
(1) a goal-oriented navigational policy that replaces passive expansion with active, intent-driven routing guided by a target-aware discriminator; 
(2) a recursive state evolution mechanism that leverages cross-request state persistence to achieve \textit{Indirectly Infinite Depth} (IID), allowing the system to logically traverse arbitrary graph depths over time without incurring linear latency growth; and 
(3) a trajectory-aware learning scheme with graph-based hard negative sampling to maintain precise alignment across the entire navigational trajectory.

Experiments on billion-scale industrial datasets demonstrate that IID-Nav significantly outperforms state-of-the-art baselines in retrieval effectiveness under latency constraints. Our results confirm that IID-Nav effectively mitigates search drift while maintaining high precision at extended depths, providing a robust and efficient solution for industrial retrieval.

\end{abstract}



\maketitle

\section{Introduction}
\label{sec:introduction}
Navigating information abundance, modern recommender systems face a foundational challenge: the retrieval of a minuscule, relevant subset from a corpus of billions of items under strict latency constraints (often <100ms). This matching stage acts as the critical bottleneck, defining the upper bound of recommendation quality. If the retrieval stage fails to capture a user's potential interest, no amount of sophisticated ranking downstream can recover it.

Currently, the retrieval paradigm is undergoing a fundamental shift from \textit{static extraction} to \textit{dynamic navigation}. Traditional item-to-item (i2i) heuristics operate primarily through discrete similarity matching, relying on pre-computed lookups to retrieve candidates associated with a user's historical behaviors. This approach essentially treats retrieval as a ``static lookup'' within a narrow, pre-defined scope. While more recent structured global indices (like TDM and NANN) have introduced discriminative models to broaden this scope, their search processes remain ``topologically constrained''. We contend that as user intents become increasingly complex and high-dimensional, retrieval must evolve from passive matching into an active, goal-oriented traversal of the item space. By navigating through iterative paths, the system can dynamically identify relevant candidates, effectively transforming retrieval into an intent-driven discovery.

Despite the success of graph-based indexing, industrial systems still face two critical limitations. The first is the \textbf{Interest Tunnel} effect. Because traditional methods rely on rigid index structures, search trajectories often become confined within local clusters of historical behaviors. This structural rigidity prevents the system from exploring deep-depth interests that lie beyond immediate behavioral neighbors, restricting candidates to narrow, well-trodden patterns. The second challenge is \textbf{Search Drift}. Most graph indices employ static global entry points that are decoupled from a user's real-time intent. Consequently, as search depth increases, the navigational trajectory tends to deviate from the true intent space. This misalignment leads to a significant degradation in recall, as the search fails to penetrate the more relevant, sparse regions of the corpus.

To overcome these constraints, we propose the theory of \textbf{Indirectly Infinite Depth (IID)}. Our core premise is that exploration depth should not be limited by the physical hops permitted within a single request's latency budget. Instead, retrieval should be a progressive process where intent is evolved across the temporal dimension. By formalizing retrieval as a stateful system, we implement a state relay mechanism that maintains search continuity across requests. This architecture ensures that even when a single interaction is limited to a few hops, the system can logically traverse arbitrary graph depths over time. This allows IID-Nav to penetrate deep-depth interest regions that were previously inaccessible to one-shot retrieval methods.

In this paper, we present \textbf{IID-Nav}, a navigational framework that integrates goal-oriented guidance with stateful evolution. Our primary contributions are as follows:
\begin{itemize}
    \item \textbf{Goal-Oriented Navigational Policy:} We replace passive neighborhood expansion with active, intent-driven routing. Guided by a target-aware discriminator, the system constructs search trajectories that precisely align with the user's current intent, transforming retrieval into active navigation.
    \item \textbf{Recursive State Evolution:} We introduce a mechanism for temporal cross-request state persistence. By caching and evolving navigational states across sessions, IID-Nav achieves IID exploration over time, effectively bypassing the physical bottlenecks that typically limit deep-depth interest discovery.
    \item \textbf{Trajectory-Aware Learning:} We implement a specialized training scheme using graph-based hard negative sampling. This ensures the model maintains precise intent alignment across the entire navigational trajectory, effectively mitigating the ``Search Drift'' problem inherent in deep-hop navigation.
    \item \textbf{Large-Scale Validation:} We provide comprehensive empirical evidence through billion-scale industrial experiments, confirming that IID-Nav significantly improves retrieval effectiveness while maintaining production-grade responsiveness.
\end{itemize}

\begin{figure*}[t]
    \centering
    \includegraphics[width=0.95\linewidth]{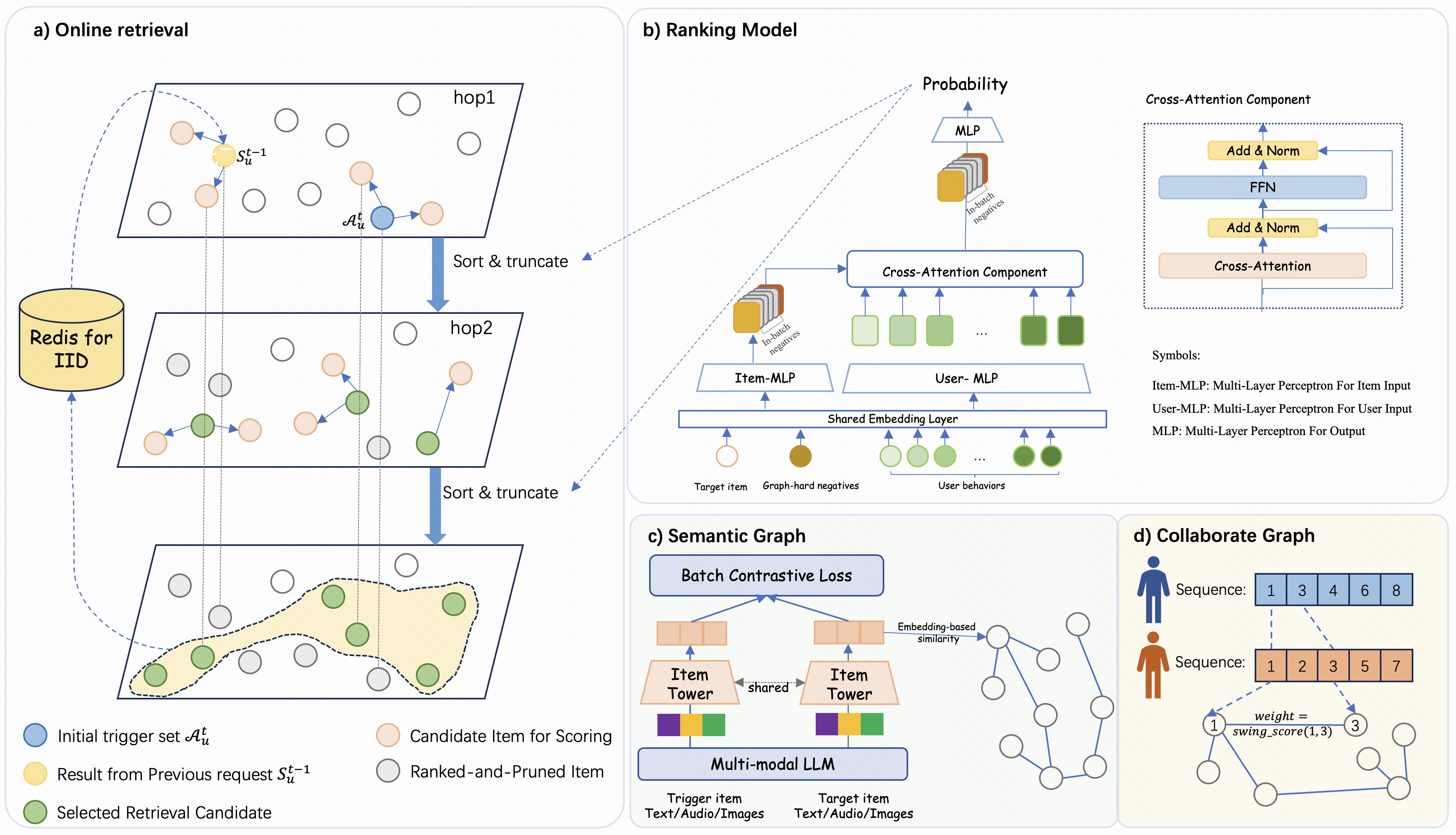}
    \caption{Overview of the IID-Nav framework. (a) Online Retrieval: Stateful Online Navigation initiates a multi-hop traversal beginning from personalized entrypoints $\mathcal{E}_u$. Search states are persisted across consecutive requests to achieve Indirectly Infinite Depth (IID) exploration within the item space. (b) Ranking Model: The target-aware discriminator directs the navigational policy through a cross-attention mechanism and is optimized with graph-hard negative sampling. (c) The semantic graph and (d) the collaborative graph constitute a heterogeneous environment. These structures are constructed via multi-modal LLMs and the Swing algorithm to provide diverse navigational trajectory.}
    \label{fig:framework}
\end{figure*}

\section{Related Work}
\label{sec:related_work}
\subsection{From Static Heuristic Extraction to Dynamic Representation}
Early industrial retrieval relied primarily on \textit{Static Heuristic Extraction}, exemplified by item-based collaborative filtering and the Swing algorithm~\cite{yang2020large}. These methods operate through a fixed matching paradigm: they pre-compute item-to-item (i2i) similarity scores and perform \textbf{static lookups} during inference. To introduce model-driven flexibility into this paradigm, Path-aware Deep Networks (PDN)~\cite{li2021path} evolve the static i2i lookup into a learned transition process. By employing a neural discriminator to score and guide structured user-item transitions, PDN captures high-order connectivity that traditional heuristic rules often overlook. This process is inherently reactive and confined by the \textbf{Interest Tunnel} effect, as it can only retrieve items with pre-existing physical links, failing to explore \textbf{deep-depth interests} beyond local behavioral clusters. 

More recently, sequence-based retrieval algorithms have gained prominence by leveraging sequential user behavior to capture dynamic preference patterns. SASRec~\cite{kang2018self} pioneered the application of self-attention in sequential recommendation, effectively modeling long-term dependencies and relative importance in sparse user-item interaction sequences without relying on manual feature engineering. Building on this foundation, MPFormer\cite{sun2025mpformer} introduced a multi-task Transformer framework to present multi-interest preferences, addressing the challenge of heterogeneous sequence patterns. In contrast, BERT4Rec~\cite{sun2019bert4rec} adopted a bidirectional transformer architecture, breaking the unidirectional constraint of traditional sequential models by masking partial items and learning contextual dependencies from both past and future interactions, thereby improving the accuracy of preference prediction. These sequence-based methods have significantly advanced the representation capability of user embeddings by integrating temporal dynamics, outperforming traditional non-sequential EBR approaches in scenarios with rich sequential behavior. 

In parallel, several studies have extended sequence-based retrieval to long-sequence settings, aiming to better capture users’ long-term preferences. ULIM~\cite{meng2025user} models ultra-long user behavior histories by disentangling them into multiple long-term interests, avoiding heuristic truncation of interaction sequences for retrieval. LongRetriever~\cite{ren2025longretriever} further studies candidate retrieval under ultra-long sequences, addressing representation degradation and efficiency challenges at industrial scale through scalable sequence encoding mechanisms. These studies indicate that effectively leveraging long and ultra-long user histories has become an emerging direction in embedding-based retrieval.

While these methods offer powerful representation-level fluidity, their candidate identification remains a one-shot event in a fixed embedding space. Despite their efficiency, such embedding-based retrieval (EBR) is fundamentally limited in expressiveness. By decomposing user-item affinity into independent embeddings, EBR cannot fully exploit deep interactive networks~\cite{zhou2018deep} or model complex feature correlations, often leading to potential mismatches with downstream rankers. This limitation necessitates a shift toward retrieval frameworks where arbitrarily complex neural models can directly govern the search process, bridging the gap between expressiveness and scalability.

\subsection{Structured Indexing and the Search Drift Challenge}
To overcome the expressiveness bottleneck of EBR, single-stage retrieval methods co-design the model and a learned index. Tree-based methods, such as TDM~\cite{2018tdm}, formulate retrieval as a beam search on a hierarchical tree scored by a deep network. Building on this framework, JOT~\cite{jtm} introduces a joint optimization strategy for tree structure and node scoring networks, reducing the misalignment between training and inference stage. MISS~\cite{guo2025miss} proposes a multi-modal index tree and a multi-modal lifelong sequence modeling module, while Deep Retrieval~\cite{dr2021} designs a multi-path hierarchical structure for item clustering. However, these tree-based approaches are fundamentally constrained by their static and rigid topologies. Once the hierarchy is constructed, the navigational path for any given intent is largely predefined by the index structure, which lacks the flexibility to adapt to real-time, heterogeneous interest shifts in a dynamic environment.

In parallel, quantization-based methods like StreamVQ~\cite{streamingvq} optimize retrieval efficiency through vector quantization and multi-stage codebook refinement. Furthermore, the GRank~\cite{sun2025grank} framework attempts to bridge retrieval and ranking by integrating a candidate generator with a lightweight ranking-aware scorer to improve target relevance.

Despite their algorithmic sophistication, these structured and quantization-based approaches frequently encounter \textbf{Search Drift}. Because they remain anchored to static global entry points or rigidly partitioned codebooks, the search trajectory is easily decoupled from the user's real-time intent. Specifically, StreamVQ's performance is often sensitive to discretization errors within its fixed partitions, while GRank’s single-shot generation mechanism lacks the flexibility to adaptively re-route its search as intent evolves. As the search penetrates deeper into these predefined topologies, the navigational process tends to deviate from the user’s true intent space, leading to a significant drop in recall coverage. Our work, \textbf{IID-Nav}, addresses this rigidity by replacing fixed hierarchies with a dynamic navigational policy that orchestrates paths based on \textbf{real-time intent alignment}.

\subsection{Graph-Based Navigation and the IID Paradigm}
Graphs have been extensively utilized in recommendation to capture complex item relationships and high-order collaborative signals. Early efforts in this domain focused primarily on graph-based representation learning, where methods such as PinSage~\cite{ying2018graph} leverage random walks and neighborhood aggregation to produce expressive item embeddings. Similarly, GNN-based models like NGCF~\cite{wang2019neural} and LightGCN~\cite{he2020lightgcn} propagate connectivity signals across user-item bipartite graphs to capture deep structural dependencies. While these methods significantly enhance the quality of the embedding space, they typically rely on traditional nearest-neighbor matching for the final retrieval, which limits their ability to actively navigate the graph topology during inference.

To bridge the gap between static representation and active search, graph-based indexing structures such as HNSW, DiskANN~\cite{jayaram2019diskann}, and NSG~\cite{fu2017fast} have been developed to enable efficient greedy traversal across large-scale item corpora. Building on these indices, Neural Approximate Nearest Neighbor (NANN)~\cite{nann} introduces neural guidance into the traversal process, treating the item space as a navigable environment for iterative search. Despite this progress, NANN and its variants operate under a stateless paradigm where the search depth is strictly capped within the computational budget of a single request. This rigid latency-depth trade-off often prevents the search process from reaching deep-depth interest, resulting in the ``interest tunnel'' effect where trajectories are confined to local, high-density behavior clusters. 

IID-Nav distinguishes itself from these single-pass navigational methods by introducing the theory of \textbf{Indirectly Infinite Depth} (IID). Unlike existing graph, tree, or quantization structures that operate within the physical constraints of a single interaction, IID-Nav implements a stateful exploration mechanism. By leveraging cross-request state persistence, our framework allows the search frontier to evolve and accumulate depth across consecutive user interactions. This evolutionary approach breaks the shackles of physical hop limits, enabling the system to uncover deep-depth interests and navigate the ``missing links'' in the global item corpus that were previously unreachable.

\begin{algorithm}[ht]
\caption{Online Multi-hop Graph Retrieval with Indirectly Infinite Depth (IID)}
\label{alg:omgr}
\begin{algorithmic}[1]
\Require 
    $A_u$: user $u$'s recent interest $\mathcal{L}_u$ and missing-memory interest $\mathcal{M}_u$\\
    $S_u$: cached recall set from last request ($|S_u| \leq 500$) \\
    $\textsc{Swing}(\cdot)$: collaborative graph expansion returning 1-hop neighbors \\
    $\textsc{Semantic}(\cdot)$: semantic graph expansion returning 1-hop neighbors \\
    $\textsc{RankModel}(\cdot; u)$: target-attention ranking model \\
    $H$: number of hops ($H=2$) \Comment{Fixed hops per request} \\
    $K$: final recall size \\
    $\Delta t$: cache expiration time
\Ensure 
    $\mathcal{S}_N$: top-$K$ recalled items for user $u$
\Statex
\Function{OnlineRetrieval}{$u$, $A_u$, $S_u$, $N$, $K$, $\Delta t$}
    \State $\mathbf{T}_1 \gets \emptyset$ \Comment{Initialize trigger set}
    \State $\mathbf{S}_u \gets \textsc{GetRedis}(\mathtt{recall:}\{u\})$ \Comment{Fetch cached IID set from previous request}
    \If{$\mathbf{S}_u \neq \emptyset$}
        \State $\mathbf{T}_1 \gets \textsc{UniqueTruncate}(\mathbf{S}_u \cup A_u, 1000)$ \Comment{IID: Chain exploration across requests}
    \Else
        \State $\mathbf{T}_1 \gets A_u$
    \EndIf

    \Statex
    \State $\mathcal{M} \gets \emptyset$ \Comment{Global score map: item $\mapsto$ score}
    
    \For{$i \gets 1$ \textbf{to} $H$}
        \State $\mathbf{C}_i \gets \textsc{Swing}(\mathbf{T}_i) \cup \textsc{Semantic}(\mathbf{T}_i)$ \Comment{Candidate expansion}
        
        \State $\mathbf{C}_i^{\text{new}} \gets \{x \in \mathbf{C}_i \mid x \notin \text{dom}(\mathcal{M})\}$ \Comment{Unscored items}
        \If{$\mathbf{C}_i^{\text{new}} \neq \emptyset$}
            \State $\mathbf{p}_{\text{new}} \gets \textsc{RankModel}(\mathbf{C}_i^{\text{new}}; u)$ \Comment{Score new items}
            \State Update $\mathcal{M}[x] \gets \mathbf{p}_{\text{new}}[x]$ for each $x \in \mathbf{C}_i^{\text{new}}$
        \EndIf
        
        \State $\mathbf{p}_i \gets [\mathcal{M}[x] \text{ for } x \in \mathbf{C}_i]$ \Comment{Retrieve scores}
        \State $\mathbf{R}_i \gets \textsc{Top-$K$}(\mathbf{C}_i, \mathbf{p}_i)$ \Comment{Select top-K}
        \State $\mathbf{T}_{i+1} \gets \mathbf{R}_i$ \Comment{Update trigger}
    \EndFor
    \Statex
    
    \State $\mathcal{R}_N \gets \mathbf{R}_N$ \Comment{Final results}
    \State $\textsc{SetRedis}(\texttt{recall:}u, \mathcal{R}_N, \Delta t)$ \Comment{Cache for IID in next request}
    \State \Return $\mathcal{R}_N$
\EndFunction
\end{algorithmic}
\end{algorithm}

\section{Methodology}
\label{sec:methodology}
As illustrated in Figure~\ref{fig:framework}, the \textbf{IID-Nav} framework consists of an offline environment construction stage and an online stateful navigation process. It reconciles high-precision discriminative scoring with billion-scale exploration through a decoupled architecture.

\subsection{Heterogeneous Navigational Environment}
\label{subsec:environment}
The navigational environment is built upon two complementary graph structures, providing the structural foundation for intent-driven exploration as illustrated in Figure~\ref{fig:framework}c, d.

\noindent \textbf{Collaborative Graph ($G_c$):} As shown in Figure~\ref{fig:framework}d, $G_c$ captures behavior-level item correlations. We utilize the Swing algorithm to compute similarity scores between items $i$ and $j$ based on their shared user interaction sequences. The edge weight $w_{ij}$ is defined by the overlap of user sets, ensuring that the navigation follows paths with high collaborative relevance.

\noindent \textbf{Semantic Graph ($G_s$).} To bridge behavioral gaps, Figure~\ref{fig:framework}c illustrates the construction of $G_s$. We employ a multi-modal Large Language Model (LLM) to extract features from item metadata (text, audio, images). Two shared-weight Item Towers are trained via batch contrastive loss to map diverse modalities into a unified embedding space. Edges are then established based on cosine similarity, enabling the system to navigate toward semantically related candidates even in the absence of co-occurrence data.

\subsection{Goal-Oriented Navigational Policy}
\label{subsec:policy}
The navigational policy is governed by a target-aware ranking model (Figure~\ref{fig:framework}b) that acts as a discriminator to guide the search frontier.

\noindent \textbf{Ranking Model Architecture.} As depicted in the right panel of Figure~\ref{fig:framework},  we employ a deep network with target-aware attention to capture fine-grained user-item interactions. Given a user $u$ with historical behavior sequence $S_u = [i_1, i_2, ..., i_L]$ and a target item $i_t$, the model first encodes both historical items and the target into dense representations. A multi-head target-attention mechanism aggregates historical behaviors conditioned on the target item:
\begin{equation}
    \text{Attention}(Q, K, V) = \text{softmax}\left(\frac{QK^T}{\sqrt{d_k}} + M\right)V,
\end{equation}
where $Q$ denotes the target item embedding, $K$, $V$ correspond to historical item embeddings, and $M$ is a mask embedding used to exclude invalid or padded historical items from the attention computation. The output is a unified user representation $z_u$ conditioned on the target $i_t$. This representation is then fed through a Multi-Layer Perceptron (MLP) alongside target item features to produce the final matching score $f_\theta(u, i_t)$. Crucially, the model architecture is \textit{not constrained} to a two-tower structure, allowing free use of cross-features and deep interactions between user and item representations, enabling the model to directly orchestrate paths toward high-utility regions.

\subsection{Trajectory-Aware Learning}
To prevent the navigational process from deviating into irrelevant item spaces, a phenomenon referenced as \textit{search drift},, we design a trajectory-aware learning scheme. To align the training objective with the multi-hop navigational paradigm, we propose a decoupled training framework with \textbf{Graph-Hard Negative Sampling}. For each positive user–item interaction $(u, i^+)$, we construct a negative set $N$ consisting of two complementary sources: (1) in-batch negative samples, and (2) graph-hard negatives sampled from the retrieval graphs ($G_c$ and $G_s$), aligning the training objective with local decision boundaries encountered during online navigation.

\noindent \textbf{Hard Negative Strategy:} We first randomly sample a candidate subset from the neighbors of $i^+$ in the retrieval graphs, capturing items that are topologically close in $G_c$ or $G_s$ but lack positive interaction labels. From this subset, we select items with \textit{lower} graph edge strength despite being topologically proximate. The intuition is that these neighbors are structurally similar but intent-dissimilar, they are the primary cause of \textbf{Search Drift} during multi-hop traversal. By forcing the discriminator to distinguish $i^+$ from these ambiguous neighbors, we ensure the model remains anchored to the true intent space as navigation depth increases.

To jointly model absolute relevance and relative ranking, the training objective combines an InfoNCE loss with a margin-based pairwise ranking loss:
\begin{equation}
    \mathcal{L} = \lambda_0\mathcal{L}_{\text{InfoNCE}} + \lambda_1\mathcal{L}_{\text{PW}},
\end{equation}
where $\lambda_0$ and $\lambda_1$ denote the weights of the InfoNCE loss and the pairwise loss respectively. 

The InfoNCE loss encourages the model to distinguish positive interactions from negative ones through contrastive learning:
\begin{equation}
\mathcal{L}_{\text{InfoNCE}}
= - \log
\frac{\exp\left(f_\theta(u, i^+) / \tau \right)}
{\exp\left(f_\theta(u, i^+) / \tau \right)
+ \sum_{i^- \in \mathcal{N}}
\exp\left(f_\theta(u, i^-) / \tau \right)},
\end{equation}
while the pairwise loss enforces a margin between the positive item and each negative:
\begin{equation}
    \mathcal{L}_{\text{PW}}
        = \sum_{i^- \in N}
        \max \bigl( 0,\;
        f_\theta(u, i^-) - f_\theta(u, i^+) + m(i^-) \bigr) .
\end{equation}
Here, the margin $m(i^-)$ is conditioned on the negative type, enabling the model to learn differentiated decision boundaries for random negatives and graph-hard negatives. This design encourages stronger separation for structurally similar but weakly connected items, while maintaining robustness against globally irrelevant negatives. 

By integrating graph-aware negative construction with a hybrid pointwise–pairwise objective, this strategy homogenizes the training signal with the iterative decision-making process of IID-Nav. Unlike prior methods that require complex adversarial alignment, IID-Nav achieves trajectory consistency by design: the discriminator is explicitly optimized to resist the cumulative errors of multi-hop traversal. As a result, even as the system achieves Indirectly Infinite Depth across requests, the navigational policy remains anchored to the user's true intent space, effectively bridging the gap between deep-depth exploration and high-precision retrieval.

\subsection{Dynamic Anchor Awakening: Intent-Driven Initialization}
\label{subsec:awakening}
Efficient large-scale retrieval critically depends on initiating the navigation from regions of the item space that are already highly aligned with the user’s current interests, rather than from fixed global entry point. To this end, IID-Nav implements a \textbf{Dynamic Anchor Awakening} mechanism to construct a personalized entrypoint set $\mathcal{E}_u^t$ for each user request at time $t$, serving as the initial trigger set. 

To comprehensively capture the user's multi-interests while mitigating the risk of ``interest narrowing'', the initial trigger set is formed by fusing two complementary signals:
\begin{itemize}[leftmargin=*, noitemsep]
    \item \textbf{Recent Interest ($\mathcal{L}_u$)}: We extract the $L$ most recent unique items from user's exposure history $\mathcal{S}_u$, sorted chronologically. This set captures persistent and up-to-date preferences, providing a high-precision starting point for local exploitation.
    \item \textbf{Missing-memory Interest ($\mathcal{M}_u$)}: To counteract \textit{recency bias}, we recover earlier historical items whose category tags are underrepresented in $\mathcal{L}_u$ based on a frequency threshold $\tau$. This reactivates latent interests from coherent but recently neglected categories.
\end{itemize}
The final initialized frontier is defined as the union $\mathcal{A}_u^t = Truncate(\mathcal{L}_u \cup \mathcal{M}_u$), ensuring the navigation starts from a diversified seed that balances temporal recency with the coverage of deep, long-tail interests. By mapping these asynchronous signals into the heterogeneous graph environment, the system ensures that the subsequent navigational policy begins its search within the most relevant subspaces of the item corpus.

\subsection{Recursive State Evolution for IID}
\label{subsec:online_iid}
The online execution follows a stateful, multi-hop traversal process (Figure~\ref{fig:framework}a) that reconciles efficiency with long-term exploration.

\noindent \textbf{Model-Guided Exploration:} The process proceeds for $H$ hops. At each hop $k$, the system expands neighbors from $G_c$ and $G_s$. The discriminator $f_\theta(u, \cdot)$ scores these candidates to identify the top-$K$ items for the next hop. This \textit{expand-score-truncate} cycle ensures active steering by real-time intent.

\noindent \textbf{Indirectly Infinite Depth (IID).}To overcome the physical hop constraints and latency budgets of single-shot retrieval, we introduce the theory of \textbf{Indirectly Infinite Depth (IID)}. We define a Navigational State $\mathcal{S}_t$ at request $t$, representing the exploration progress reached during that interaction. The transition between consecutive requests is defined as:\begin{equation}\mathcal{E}{u}^{(t+1)} = \mathcal{A}{u}^{(t+1)} \cup \mathcal{S}_u^t,
\end{equation}
where $\mathcal{A}{u}^{(t+1)}$ denotes the newly awakened anchors and $\mathcal{S}_u^t$ injects high-quality terminal nodes from the previous request's search frontier into the current initial navigation points. As shown in Figure~\ref{fig:framework}a, this asynchronous state handoff is implemented via a Redis-based relay. By caching and recycling high-utility nodes, IID-Nav allows the exploration depth to evolve across the temporal axis. This mechanism enables the system to logically reach an arbitrary depth in the item subspace over multiple interactions, effectively penetrating sparse interest regions that are unreachable within the hop limit of any single request.

\{input\begin{table*}[t]
\centering
\caption{Performance Comparison of SOTA Methods on Different Datasets. The best and second-best results highlighted in bold font and \underline{underlined}. \textit{Note: TDM was only evaluated on the Industry Dataset due to its deployment complexity, using our existing infrastructure.} }
\label{tab:baseline_cmp}
\begin{tabular}{lccccccccccc}
\toprule
\multirow{2}{*}{Method} & \multicolumn{2}{c}{User Behavior} & \multicolumn{2}{c}{MovieLens} & \multicolumn{3}{c}{Industry Datasets} \\
\cmidrule(lr){2-3} \cmidrule(lr){4-5} \cmidrule(lr){6-8}
& Recall@50 & NDCG@50  & Recall@50 & NDCG@50 & Recall@500 & NDCG@500 & QPS\\
\midrule
DSSM  & 0.2711 & 0.1911 & 0.1940 & 0.0792 & 0.1068 & 0.0360  & \textbf{1300}\\
Kuaiformer  & \underline{0.3270} & \textbf{0.2347} & 0.2538 & 0.0906 & 0.1151 & 0.0386 & 772\\
TDM & - & - & - & - & \underline{0.1766} & 0.0535 & 273\\
NANN & 0.3264 & 0.1765 & \underline{0.2633} & \underline{0.1017} & 0.1326 & 0.0466 & 384\\
Streaming VQ & 0.3220 & \underline{0.2262} & 0.2554 & 0.0907 & 0.1296 & \underline{0.0603} & 450 \\

\midrule
IID-Nav  & \textbf{0.4920} & 0.2256  & \textbf{0.3642} & \textbf{0.1304} & \textbf{0.2408} & \textbf{0.0826}   &  \underline{910} \\
Relative Gain & +50.46\% & -3.88\% & +38.32\% & +28.22\% & +36.35\% & +36.98\% & -30.0\%\\

\bottomrule
\end{tabular}
\end{table*}

\section{Experiments}
\label{sec:experiment}
In this section, we evaluate \textbf{IID-Nav} through extensive offline experiments and large-scale online A/B tests. We aim to verify whether the stateful navigation mechanism can effectively overcome the depth constraints of traditional retrieval and resolve the \textit{Search Drift} challenge via trajectory-aware learning.

\subsection{Research Questions}
\label{subsec:rqs}
Our experiments are designed to address the following research questions (RQs):
\begin{itemize}[leftmargin=*, noitemsep]
\item \textbf{RQ1:} Does IID-Nav outperform state-of-the-art retrieval methods in terms of precision and coverage?
\item \textbf{RQ2:} How does the IID mechanism affect the exploration depth and interest discovery over time?
\item \textbf{RQ3:} Can the trajectory-aware learning effectively mitigate ``Search Drift'' during multi-hop navigation?
\item \textbf{RQ4:} How do the multi-graph fusion and anchor awakening contribute to the performance?
\item \textbf{RQ5:} Is the framework efficient and robust enough for large-scale industrial deployment?
\end{itemize}

\subsection{Experimental Settings}
\label{subsec:settings}

\noindent \textbf{Datasets.} We evaluate IID-Nav on one industrial and two public datasets (Table~\ref{tab:dataset_stats}): 
\textbf{MovieLens-20M} (movie recommendations), \textbf{Taobao UserBehavior} (e-commerce), and \textbf{ShortVideo-Ind} (a representative short-video platform). All datasets are split chronologically. For the ShortVideo-Ind dataset, we utilize six consecutive days for training and the subsequent day for testing. The public datasets follow a standard split, with the initial 90\% of logs used for training and the remaining 10\% for evaluation.

\begin{table}[h]
\centering
\caption{Dataset statistics after preprocessing}
\label{tab:dataset_stats}
\begin{tabular}{lccc}
\toprule
 & {UserBehavior} &{MovieLens-20M} & {ShortVideo-Ind} \\
\midrule
Users & 964K & 138K & 108M \\
Items & 4.2M & 27K & 42M  \\
Interactions & 1.7M & 9.3M & 4B \\
Avg. Seq. Len. & 101 & 144 & 980 \\
\bottomrule
\end{tabular}
\vspace{-1em}
\end{table}

\noindent \textbf{Baselines.} We compare IID-Nav against several representative baselines:
\begin{itemize}[leftmargin=*, noitemsep]
    \item \textbf{DSSM~\cite{huang2013learning}:} A standard two-tower Deep Neural Network (DNN) model. User and item embeddings are learned via inner product, and retrieval is performed using the FAISS library with an HNSW index.
    \item \textbf{Kuaiformer~\cite{kuaiformer}:} A Transformer-based two-tower retrieval model that enhances user representation via sequence modeling. User and item embeddings are learned independently and matched by inner product, with retrieval conducted using FAISS and an HNSW index.
    \item \textbf{TDM~\cite{2018tdm}:} The tree-based deep retrieval model. We implement this baseline using a deep matching network comparable in capacity to our discriminator to ensure a fair comparison. Unlike our dynamic navigational policy, TDM performs retrieval via beam search over a fixed hierarchical tree.
    \item \textbf{NANN~\cite{chen2022nann}:} A graph-based retrieval method utilizing model-guided traversal. We construct its index using item embeddings ($l2$ distance) and incorporate its specific adversarial training task. Notably, NANN operates in a stateless paradigm, with its exploration depth strictly confined to the computational budget of a single request.
    \item \textbf{Streaming VQ~\cite{streamingvq}:} A method combining vector quantization with graph traversal. Similar to other stateless baselines, its navigational range is constrained by the single-pass search requirement.
\end{itemize}
\vspace{0.5em}

\noindent \textbf{Evaluation Metrics.}
For offline evaluation, we adopt \textbf{Recall@K} and \textbf{NDCG@K} (K=50, 500),  alongside \textbf{Queries Per Second (QPS)} for system efficiency  under a stringent \textbf{100ms latency constraint} to ensure production-grade responsiveness. For the online A/B test, we monitor key business metrics: Total App Usage Time, Usage Time per User, and Video Watch Time. 
\vspace{0.5em}

\noindent \textbf{Implementation Details.}
IID-Nav utilizes a target-attention architecture with multi-head attention. 
We construct two item–item graphs to support retrieval: (1) a collaborative graph based on Swing-style co-interaction signals, and (2) a semantic graph based on representation-level content similarity. For each trigger item, we retain at most 50 neighboring items in each graph to control graph density and ensure efficient traversal.
During online retrieval, IID-Nav dynamically generates a personalized entrypoint set $\mathcal{E}_u$ with a maximum size of 1,000 items. Starting from this set, the retrieval process performs $H=2$ hops of graph exploration. At each hop, candidate expansion is bounded such that the total number of evaluated items is capped at approximately 15,000 items per hop, balancing retrieval coverage and latency constraints.
The model is trained using the Adam optimizer with standard hyperparameter settings. All reported configurations are chosen to meet strict online serving requirements while preserving sufficient exploration capacity for complex, model-guided graph retrieval.

\subsection{Overall Performance (RQ1)}
\label{subsec:overall}

The overall performance comparison is presented in Table~\ref{tab:baseline_cmp}. IID-Nav consistently and significantly outperforms all baseline methods in terms of Recall@K across all datasets, demonstrating its superior ability to discover relevant items through model-guided exploration. 

Specifically, on the MovieLens and Industrial datasets, IID-Nav achieves the best results in both Recall and NDCG. On the User Behavior dataset, while NDCG is slightly lower than Kuaiformer, IID-Nav prioritizes a massive Recall gain (+50.46\%) to provide a more comprehensive candidate pool. This trade-off is intentional: in industrial multi-stage pipelines, a higher recall rate is more critical for overall system performance than minor ranking variations at the retrieval stage. By leveraging an expressive cross-attention discriminator to guide paths across heterogeneous graphs, IID-Nav effectively penetrates sparse interest regions that traditional two-tower or tree-based models fail to reach.

\subsection{Deep Exploration Analysis (RQ2 \& RQ3)}

\begin{figure}[htbp]
    \centering
    \includegraphics[width=0.85\linewidth]{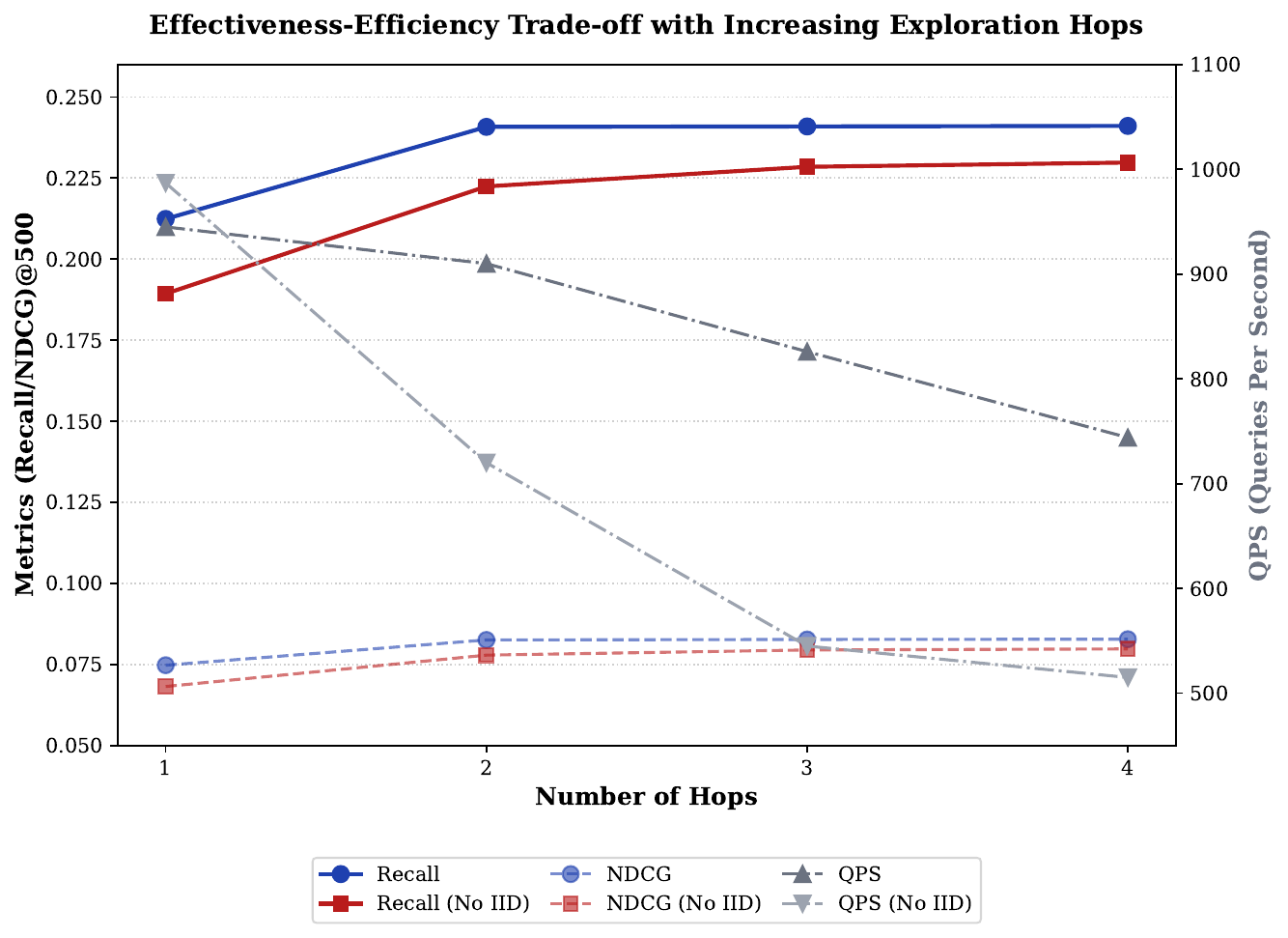}
    \caption{Recall@500 and QPS versus the number of exploration hops ($H$). Performance plateaus after 2 hops while QPS declines linearly. The infinite exploration variant outperforms all fixed-hop settings, approaching the asymptotic behavior of infinite hops.}
    \label{fig:hops}
\end{figure}

\noindent \textbf{Analysis of Multi-hop Exploration and IID.}
We evaluate the multi-hop mechanism and its \textit{Indirectly Infinite Depth (IID)} variant to determine the optimal configuration. Figure~\ref{fig:hops} illustrates the trade-off between Recall@500 and QPS relative to the number of hops ($H$).

The key observations are as follows:
1) Performance jumps significantly from 1-hop (i2i) to 2 hops, confirming the need for multi-step reasoning.
2) Gains saturate after 2 hops, while QPS degrades linearly with $H$.
3) The \textit{IID mechanism} consistently outperforms all fixed-hop configurations. Notably, the configuration with $H=2$ combined with IID exploration achieves the best performance, surpassing even the results of fixed-hop models with $H > 4$. This confirms that IID successfully approximates the behavior of infinite-hop exploration by accumulating depth across requests without increasing per-request latency.

Thus, we select \textbf{H=2 with IID} as the optimal point, balancing effectiveness and efficiency. This is the default in our experiments.

\vspace{0.5em}
\noindent \textbf{Robustness against Search Drift.}
In graph-based navigation, an inherent risk is \textit{Search Drift}, where the expansion process deviates from the user's core intent as depth increases. While Figure~\ref{fig:hops} shows that Recall grows with more hops, we contend that the precision of this expansion is governed by the discriminator’s ability to filter out topologically proximate but semantically irrelevant neighbors.

Our ablation results in Table~\ref{tab:ablation_graph} provide empirical evidence for this: removing \textbf{graph-hard negatives} from the training objective leads to a significant \textbf{8.26\%} drop in Recall@500. This decline indicates that without being explicitly trained to distinguish the target from topologically proximate ``hard'' distractors in $G_c$ and $G_s$, the navigational policy becomes susceptible to drift. By incorporating these challenging negatives, \textbf{IID-Nav} anchors each step of the multi-hop traversal to the user's true intent. This anchoring remains effective even during state-relay across requests, successfully suppressing the accumulation of navigational errors. Consequently, the framework maintains high precision throughout deep-depth exploration.

\subsection{Ablation Study (RQ4)}
\begin{table}[h]
    \centering
    \caption{Ablation study on the graph.}
    \label{tab:ablation_graph}
    \begin{tabular}{lcc}
        \toprule
        {Variant} & {Recall@500} & {NDCG@500} \\
        \midrule
        \textbf{IID-Nav (Full)} & \textbf{0.2408} & \textbf{0.0826} \\
        \midrule
        only CF graph & 0.1852 & 0.0664 \\
        only Semantic graph & 0.1546 & 0.0547 \\
        w/o graph-hard negatives & 0.2209 & 0.0755 \\
        \bottomrule
    \end{tabular}
\end{table} 

\noindent\textbf{Multi-Graph Fusion.} We evaluate the impact of heterogeneous graph design. As shown in Table~\ref{tab:ablation_graph}, using only the Collaborative Graph yields high precision but limited diversity. Conversely, using only the Semantic Graph improves novelty but hurts immediate relevance. Their fusion achieves the optimal balance between exploitation and exploration.

\begin{table}[h]
    \centering
    \caption{Ablation study on the anchor mechanism.}
    \label{tab:ablation_entry}
    \begin{tabular}{lcc}
        \toprule
        \textbf{Variant} & \textbf{Recall@500} & \textbf{NDCG@500} \\
        \midrule
        \textbf{IID-Nav (Full)} & \textbf{0.2408} & \textbf{0.0826} \\
        \midrule
        w/o recent ($\mathcal{M}_u$) & 0.1388 & 0.0434 \\
        w/o missing-memory ($\mathcal{L}_u$) & 0.1898 & 0.0605 \\
        \midrule
        IID-Nav-Random & 0.0980 & 0.0315 \\
        IID-Nav-Popular & 0.1264 & 0.0401 \\
        \bottomrule
    \end{tabular}
\end{table}

\noindent\textbf{Anchor Strategies.} As shown in Table~\ref{tab:ablation_entry}, the \textbf{IID-Nav (Full)} model consistently achieves the highest performance across all metrics. It significantly outperforms the Random selection and most popular anchor strategies by 145.7\% and 90.5\% in Recall@500 respectively. These results indicate that using static global triggers as starting points often leads to sub-optimal navigation due to the lack of user-specific intent localization. Furthermore, removing the missing-memory interest ($\mathcal{M}_u$) results in a 21.2\% drop in Recall. This degradation confirms that reactivating historical anchors is essential for mitigating the ``interest tunnel'' effect and exploring regions beyond local behavioral clusters. The integration between recent interactions ($\mathcal{L}_u$) and long-term memory ensures that the navigational process begins from a high-relevance intent space.

\subsection{Efficiency and Online A/B Test (RQ5)}
\label{subsec:efficiency}

\noindent \textbf{Efficiency Benchmark.} Table~\ref{tab:baseline_cmp} shows the online serving efficiency of IID-Nav. Benefiting from targeted search via personalized anchors, IID-Nav achieves a QPS of 910, which is competitive with the highly-optimized EBR system (1300) and significantly higher than TDM (273) and NANN (384).

\noindent \textbf{Online A/B Test.} We conducted a one-week online A/B test on large-scale short-video platform (\textbf{ShortVideo-Ind}) by integrating IID-Nav as a supplementary retrieval component. As shown in Table~\ref{tab:peg_online}, IID-Nav consistently improves user engagement across different application scenarios. Specifically, in both the Main-App and Lite-App interfaces, IID-Nav yielded a 0.3\%–0.4\% gain in total application usage time and up to a 0.26\% increase in video watch time. These results demonstrate that IID-Nav effectively enhances user engagement across both standard and lightweight application settings, validating the robustness and practical value of stateful graph retrieval in large-scale production systems.

\begin{table}[ht]
\centering
\caption{Online A/B Test Results of IID-Nav Across Different Scenarios}
\label{tab:peg_online}
\begin{tabular}{lccccccc}
\toprule
Applications & \makecell{Total App \\ Usage Time} & \makecell{Usage Time \\ per User} & \makecell{Video \\ Watch Time} \\
\midrule
KS Single Page  & +0.334\% & +0.339\% & +0.553\% \\
KS Lite Page  & +0.400\% & +0.408\% & +0.566\% \\
\bottomrule
\end{tabular}
\end{table}

\section{Conclusion}
\label{sec:conclusion}
In this paper, we address the fundamental limitation of retrieval depth in large-scale recommender systems. Most existing frameworks are constrained by a rigid latency-depth trade-off within the boundaries of a single request. This constraint often prevents the exploration of deep-depth interests and sparse regions within the global item corpus. To overcome this limitation, we introduce \textbf{IID-Nav}, a framework that achieves \textbf{Indirectly Infinite Depth (IID)} through a stateful navigational paradigm.

The framework implements a goal-oriented navigational policy. This policy replaces passive neighborhood expansion with active, intent-driven routing directed by a target-aware discriminator. Furthermore, the recursive state evolution mechanism leverages cross-request state persistence to achieve IID exploration. This approach allows the system to logically traverse arbitrary distances in the item graph without incurring linear growth in per-request latency. To mitigate the risk of \textbf{Search Drift}, we integrate a trajectory-aware learning scheme with graph-based hard negative sampling. This scheme ensures precise intent alignment across the entire navigational trajectory. By anchoring the search frontier to the user’s true intent, IID-Nav successfully suppresses the accumulation of errors that typically occurs during deep-depth exploration.

Extensive evaluations on public and billion-scale industrial datasets confirm the effectiveness of IID-Nav. Our results demonstrate that decoupling search depth from single-request physical limits significantly improves retrieval coverage and user engagement. By providing a scalable solution that maintains high precision at extended depths, IID-Nav establishes stateful, autonomous navigation as a robust and practical direction for next-generation retrieval systems.

\bibliographystyle{ACM-Reference-Format}
\bibliography{sample-base}

\end{document}